\documentclass[11pt] {article}

\usepackage{epsfig}
\catcode`@=11

\def\@citex[#1]#2{\if@filesw\immediate\write\@auxout{\string\citation{#2}}\fi
  \def\@citea{}\@cite{\@for\@citeb:=#2\do
    {\@citea\def\@citea{,\penalty\@m}\@ifundefined
      {b@\@citeb}{{\bf ?}\@warning
       {Citation `\@citeb' on page \thepage \space undefined}}%
\hbox{\csname b@\@citeb\endcsname}}}{#1}}

\def\citer{\@ifnextchar [{\@tempswatrue\@citexr}{\@tempswafalse\@citexr[]}}

\def\@citexr[#1]#2{\if@filesw\immediate\write\@auxout{\string\citation{#2}}\fi
  \def\@citea{}\@cite{\@for\@citeb:=#2\do
    {\@citea\def\@citea{--\penalty\@m}\@ifundefined
       {b@\@citeb}{{\bf ?}\@warning
       {Citation `\@citeb' on page \thepage \space undefined}}%
\hbox{\csname b@\@citeb\endcsname}}}{#1}}
\catcode`@=12
\def\s{\hat{s}}
\def\u{\hat{u}}
\def\z{v \cdot \hat{q}}

\def\ml{\hat{m}_l}
\def\ms{\hat{m}_s}

\def\mc{\hat{m}_c}

\def\lo{\hat{\lambda}_1}
\def\lt{\hat{\lambda}_2}

\def\q{\hat{q}}
\def\bxsll{$B \rightarrow X_s \ell^+ \ell^- $}
\def\absvcb{\left| V_{cb} \right|}

\title{  { \bf
Hadron Spectra and Spectral Moments in the Decay
 \bxsll ~using HQET}}
\author{\vspace{1cm}\\
   {\bf A.~Ali\thanks{E-mail address: ali@x4u2.desy.de} ~and 
G.~Hiller\thanks{E-mail address: ghiller@x4u2.desy.de}}\\
         Deutsches Elektronen-Synchrotron DESY, Hamburg \\
        \vspace{5mm}\\}
\date{}

\begin{document}
\setlength{\baselineskip}{24pt}

\maketitle
\begin{picture}(0,0)
       \put(325,260){DESY 98-025}
       \put(325,245){March 1998}
\end{picture}
\vspace{-24pt}
\setlength{\baselineskip}{7mm}

\begin{abstract}
We compute the leading order (in $\alpha_s$) perturbative QCD and power 
($1/m_b^2)$ corrections 
to the hadronic invariant mass and hadron energy spectra in the decay $B 
\to X_s \ell^+ \ell^-$ in standard model using the heavy quark expansion 
technique (HQET).
Results for the
first two hadronic moments $\langle S_H^n\rangle$ and $\langle 
E_H^n\rangle$, $n=1,2$, are presented here working out their sensitivity on 
the HQET parameters $\lambda_1$ and $\bar{\Lambda}$. 
Data from the forthcoming B facilities can be 
used to measure the short-distance contribution in \bxsll and   
determine the HQET parameters from the moments $\langle S_H^n\rangle$.
This can be combined with the analysis of semileptonic decays $B \to X \ell 
\nu_\ell$ to determine them precisely.

\vspace*{2.5cm}
\centerline{PACS numbers: 12.38.Bx, 13.20.He, 12.39.Hg}
\vspace*{0.5cm}
\centerline{(Submitted to Physical Review Letters)}

\end{abstract}

\thispagestyle{empty}
\newpage
\setcounter{page}{1}

  The semileptonic inclusive decays $B \to X_s \ell^+ \ell^-$ , where
$\ell^\pm = e^\pm,\mu^\pm,\tau^\pm$, offer, together with the
radiative electromagnetic penguin decay $B \to X_s + \gamma$, presently the
most popular testing grounds for the standard model (SM) in the flavor
sector. Concentrating on the decay $B \to X_s \ell^+ \ell^-$, we recall that
the first theoretical calculations including partial leading order
QCD corrections were reported a decade ago in \citer{HWS87,JW90},
 emphasizing the
sensitivity of the dilepton mass spectrum and decay rates to the top quark
mass in the short-distance contribution. 
Since the pioneering papers \citer{HWS87,JW90}, a lot of
theoretical work has been done on the decay \bxsll. 
In particular, the complete leading order 
perturbative corrections in the QCD coupling constant $\alpha_s$ 
to the dilepton invariant mass spectrum 
and so-called forward-backward (FB) asymmetry of the 
leptons \cite{amm91},
have been calculated in refs.~\cite{burasmuenz,misiakE} and \cite{AHHM97}, 
respectively. In addition, leading order power corrections in  
$1/m_b^2$ to the decay rate, dilepton invariant mass spectrum and the 
FB asymmetry have been studied 
\cite{AHHM97}, using the heavy quark expansion technique (HQET), 
correcting an earlier derivation of the dilepton
spectrum in this decay \cite{FLSold}.
 The power corrected
dilepton mass spectrum and FB asymmetry have also been derived for the 
massless  $s$-quark case recently in \cite{BI98}, confirming the  
results in \cite{AHHM97}.
 Corrections of order $1/m_c^2$
to the dilepton mass spectrum away from the resonant 
regions have also been worked out 
\cite{buchallaisidorirey,chenrupaksavage}.

In this letter, we study the hadron spectra and  moments in \bxsll using
HQET. The study of the decay $B \to X \ell \nu_\ell$ 
in this context has received a lot of interest \citer{FLS,DU97}.
 The hadronic 
invariant mass spectra in \bxsll and $B \to X_u \ell \nu_\ell$  
have 
striking similarities and differences. For example, both of these processes
have at the parton level a delta function behavior  
$d\Gamma/ds_0 \propto \delta(s_0-m_q^2)$, $q=u,s$, where $s_0$ is
the partonic invariant mass. Thus, the 
entire invariant mass spectrum away from $s_0=m_q^2$ is 
generated perturbatively (by gluon bremsstrahlung) and through  
$B$-hadron non-perturbative effects. The latter are already present
in ${\cal O}(\alpha_s^0)$, as can  be seen by  the relation 
between the $b$ quark mass and the $B$ meson mass. In HQET
this takes the form
$m_B=m_b+\bar{\Lambda} -(\lambda_1+ 3\lambda_2)/2m_b +...$. 
The quantities $\lambda_1$,
$\lambda_2$ and $\bar{\Lambda}$ are the HQET parameters 
\cite{georgi,MW,FLSold}.
Keeping, for the sake of simplicity just the $\bar{\Lambda}$ term, the
hadronic invariant mass $S_H$ is related to $s_0$ and the partonic energy
$E_0$ by $S_H=s_0 + 2\bar{\Lambda}E_0 + \bar{\Lambda}^2$.
This gives rise to a non-trivial spectrum in the entire region
$\bar{\Lambda}^2 < S_H < M_B^2$. Hence,  measurements of these
spectra would lead to direct information on the QCD dynamics
and a better determination of the non-perturbative parameters, such as
the HQET parameters $\bar{\Lambda}$ and $\lambda_1$.
Following this line of argument, the sensitivity of 
the lepton and hadron energy spectra on these parameters in the decays $B 
\to X \ell \nu_\ell$ has been studied quantitatively in literature 
\cite{FLS,gremm,FL98}; Photon energy moments in the decay $B \to X_s + 
\gamma$ have also been worked out in \cite{KL95}.
 Present status of the HQET parameters is reviewed in
\cite{neubert}. There is a fair amount of theoretical dispersion on 
$\bar{\Lambda}$ and $\lambda_1$ and it 
will be very instructive to get independent and complementary information
on these parameters from other $B$ decays. 

  We report here a calculation of the hadron 
spectra in the  decay \bxsll. 
Leading order (in $\alpha_s$) perturbative QCD and 
power ($1/m_b^2)$ corrections to the hadronic invariant mass and 
hadron energy spectra in this decay are computed at the parton level.
Including both the ${\cal O} (1/m_b^2)$
and ${\cal O}(\alpha_s)$ terms generates hadron spectra
with contributions  of ${\cal O}(\bar{\Lambda}/m_B)$,
${\cal O}(\alpha_s\bar{\Lambda}/m_B)$, ${\cal O}(\lambda_1/m_B^2)$ 
and ${\cal O}(\lambda_2/m_B^2)$. Relegating the detailed hadronic profile to
a subsequent publication \cite{AH98-2}, here
the power- and perturbatively
corrected  hadronic spectral moments  
$\langle S_H^n\rangle$ and $\langle E_H^n\rangle$
are presented for the first two moments $n=1,2$. The former 
are sensitive to the HQET parameters
$\bar{\Lambda}$ and $\lambda_1$ and we work out this dependence
numerically, showing that these moments would provide   
an independent determination of the HQET parameters in \bxsll.
The theoretically constrained contours in the $(\bar{\Lambda}-\lambda_1)$
plane in \bxsll are compared with the corresponding one from an analysis of 
the power 
corrected lepton energy spectrum in $B \to X \ell \nu_\ell$
\cite{gremm}. We argue that 
a simultaneous analysis of the moments and spectra in \bxsll and $B \to X 
\ell \nu_\ell$ will allow to determine the HQET parameters with a high 
precision.  

We start with the definition of the kinematics of the decay at the parton 
level,
$b (p_b) \to s (p_s) (+g (p_g))+\ell^{+} (p_{+})+\ell^{-}(p_{-}) $,
where $g$ denotes a gluon from the $O(\alpha_s)$ correction. The 
corresponding kinematics at the hadron level can be written as:
$B (p_B) \to X_s (p_H)+\ell^+ (p_{+})+\ell^{-} (p_{-})$.
We define by $q$ and $s$ the momentum transfer to the lepton pair and the 
invariant 
mass of the dilepton system, respectively, $q = p_{+}+p_{-}$,
$s =q^2$; the dimensionless variables with a hat
are related to the variables with dimension by the scale $m_b$, 
the $b$-quark mass, e.g., $\s= \frac{s}{m_b^2}$, $\ms=\frac{m_s}{m_b}$ etc..
Further, we define a 4-vector $v$, which denotes the velocity of both the
$b$-quark and the $B$-meson, $p_b=m_b v$ and $p_B=m_B v$. We shall also need
a variable $u$, which is defined as
$u = -(p_b-p_+)^2+(p_b-p_{-})^2$, with the scaled variable 
$\u=u/m_b^2$ satisfying the kinematic relation
$ \u = 2 v \cdot (\hat{p}_{+}-\hat{p}_{-})$.
The hadronic invariant mass is denoted by  $S_H \equiv p_H^2$ and 
$E_H$ denotes the hadron energy in the final state. 
The corresponding quantities at parton level are the 
invariant mass $s_0$ and the scaled parton energy $x_0\equiv 
\frac{E_0}{m_b}$; without gluon bremsstrahlung this simplifies to
$s_0=m_s^2$ and $x_0$ becomes directly related to the dilepton invariant 
mass, $x_0=1/2(1-\s +\ms^2)$.
{}From momentum conservation, the following equalities hold in the $b$-quark,
equivalently $B$-meson, rest frame ($v=(1,0,0,0)$):
\begin{eqnarray}
\label{eq:kin}
x_0  &=& 1- v \cdot \q \, ,
~~\s_0 = 1 -2 v \cdot \q + \s \, ,\nonumber \\
E_H &=& m_B-v \cdot q \, ,
~~S_H = m_B^2 -2 m_B v \cdot q  + s \, \, .
\end{eqnarray}
The relation between the kinematic variables of the parton model and the 
hadronic states is, using the
HQET mass relation
$m_{B}=m_b +\bar{\Lambda}-1/2 m_b (\lambda_1+3 \lambda_2)+\dots$,
given as
\begin{eqnarray}
  E_H&=&\bar\Lambda-{\lambda_1+3\lambda_2\over2m_B}+\left(m_B-\bar
  \Lambda+
  {\lambda_1+3\lambda_2\over2m_B}\right) x_0+\dots\,, \nonumber\\
  S_H&=&m_s^2+\bar\Lambda^2+(m_B^2-2\bar\Lambda m_B+\bar\Lambda^2
  +\lambda_1+3\lambda_2)\,(\hat s_0-\hat m_s^2) \nonumber \\
  &&\qquad\qquad\mbox{}+(2\bar\Lambda
m_B-2 \bar\Lambda^2-\lambda_1-3\lambda_2) x_0
  +\dots\,,
\label{ehe0shs0}
\end{eqnarray}
where the ellipses denote terms higher order in
$1/m_b$. 
The quantity $\lambda_2$ is known precisely from the $B^* - B$
mass difference, with $\lambda_2 \simeq 0.12$ GeV$^2$. The other two
parameters are of interest here.

The effective 
Hamiltonian governing the decay \bxsll, obtained by integrating out the 
top quark and the $W^\pm$ bosons, is given as
\begin{eqnarray}\label{heffbsll}
{\cal H}_{eff}(b \to s + X, \, X=\gamma, \, \ell^{+} \ell^{-}) 
= - \frac{4 G_F}{\sqrt{2}} V_{ts}^* V_{tb}
 \left[ \sum_{i=1}^{6} C_i (\mu)  O_i 
+ C_7 (\mu) \frac{e}{16 \pi^2}
          \bar{s}_{\alpha} \sigma_{\mu \nu} (m_b R + m_s L) b_{\alpha}
                F^{\mu \nu} 
\right. \nonumber \\
 \left.
+C_8 (\mu) O_8 
+ C_9 (\mu) \frac{e^2}{16 \pi^2}\bar{s}_\alpha \gamma^{\mu} L b_\alpha
\bar{\ell} \gamma_{\mu} \ell 
+ C_{10}  \frac{e^2}{16 \pi^2} \bar{s}_\alpha \gamma^{\mu} L
b_\alpha \bar{\ell} \gamma_{\mu}\gamma_5 \ell \right] \, ,
\end{eqnarray}
where $L$ and $R$ denote chiral projections, $L(R)=1/2(1\mp \gamma_5)$,
 $V_{ij}$ are the CKM matrix elements and the
CKM unitarity has been used in factoring out the product $V_{ts}^\ast
V_{tb}$. The operator basis is taken from \cite{AHHM97}, where also the 
Four-Fermi operators $O_{1},\dots ,O_{6}$ and the chromo-magnetic 
operator $O_8$ can be seen.
Note that $O_8$ does not contribute to the decay \bxsll in the 
approximation which we use here. The $C_i(\mu)$ are the Wilson coefficients,
which depend, in general, on the renormalization scale $\mu$,
except for $C_{10}$. Their numerical values are given in Table \ref{wilson}. 
\begin{table}[h]
        \begin{center}
        \begin{tabular}{|c|c|c|c|c|c|c|c|c|}
        \hline
        \multicolumn{1}{|c|}{ $C_1$}       &
        \multicolumn{1}{|c|}{ $C_2$}       &
        \multicolumn{1}{|c|}{ $C_3$}       &
        \multicolumn{1}{|c|}{ $C_4$}       &
        \multicolumn{1}{|c|}{ $C_5$}       &
        \multicolumn{1}{|c|}{ $C_6$}       &
        \multicolumn{1}{|c|}{ $C_7^{\mbox{eff}}$}       &
        \multicolumn{1}{|c|}{ $C_9$}       &
                \multicolumn{1}{|c|}{$C_{10}$} \\
        \hline
        $-0.240$ & $+1.103$ & $+0.011$ & $-0.025$ & $+0.007$ & $-0.030$ &
   $-0.311$ &   $+4.153$ &    $-4.546$      \\
        \hline
        \end{tabular}
        \end{center}
\caption{ \it Values of the Wilson coefficients used in the numerical
          calculations corresponding to the central values
          of the parameters given in eq.~(\protect\ref{eq:param}).
Here, $C_7^{\mbox{eff}} \equiv C_7 - C_5/3 -C_6$, and
for $C_9$ we use the NDR scheme.}
\label{wilson}
\end{table}   

With the help of the above expressions, one can express the 
Dalitz distribution in \bxsll as:  
\begin{equation}
        \frac{{\rm d} \Gamma}{{\rm d}\u \, {\rm d}\s \, {\rm d}(\z)} = 
                \frac{1}{2 \, m_B}
                \frac{{G_F}^2 \, \alpha^2}{2 \, \pi^2} 
                \frac{{m_b}^4}{256 \, \pi^4}
                \left| V_{ts}^\ast V_{tb} \right|^2 
                \, 2 \, {\rm Im} 
                \left( {T^L}_{\mu \nu} \, {L^L}^{\mu \nu}
                +  {T^R}_{\mu \nu} \, {L^R}^{\mu \nu} \right) \, ,
        \label{eqn:dgds}
\end{equation}
where the hadronic and leptonic tensors ${T^{L/R}}_{\mu \nu}$ and
${L^{L/R}}^{\mu \nu}$ are given in \cite{AHHM97}. 
Using Lorentz decomposition, the tensor $T_{\mu \nu}$ can be expanded in 
terms of three structure functions $T_i$,
\begin{equation}
        T_{\mu \nu} = -T_1 \, g_{\mu \nu} + T_2 \, v_\mu \, v_\nu 
                + T_3 \, i \epsilon_{\mu \nu \alpha \beta} \, 
                        v^\alpha \, \hat{q}^\beta \, ,
\label{eq:hadrontensor}
\end{equation}
where the ones which do not contribute to the
amplitude in the limit of massless leptons have been neglected.

Concerning the $O(\alpha_s)$ corrections to the hadron spectra, we note that 
only $O_9 =e^2/(16 \pi^2)\bar{s}_\alpha \gamma^{\mu} L b_\alpha 
\bar{\ell} \gamma_{\mu} \ell$ is subject to such corrections. These can be 
obtained by using the existing results in the literature as follows:
The vector current $O_9$ can be decomposed as
$V=(V-A)/2 + (V+A)/2$. Note that the $(V-A)$ and $(V+A)$ currents yield 
the same  hadron energy spectrum \cite{aliold}
and there is no interference term  present in this spectrum for massless 
leptons. So, the correction for the vector current case 
can be taken from the corresponding result for the charged $(V-A)$ case
\cite{aliqcd,jezkuhn}.

We have calculated the  order $\alpha_s$  perturbative QCD correction
for the hadronic invariant mass in the range
$\ms^2 < \s_0 \leq 1 $. Since the decay $b \to s + \ell^+ + \ell^-$
contributes in the parton model only at $\s_0 =\ms^2$,  only 
the bremsstrahlung graphs $b \to s + g + \ell^+ + \ell^-$ contribute in this 
range. This makes the calculation much simpler than in the full $\s_0$ range 
including virtual gluon diagrams. Also for this distribution, the results
can be taken from the existing literature.
As the starting point, we use  
the Sudakov exponentiated double differential 
decay rate $\frac{{\rm d^2}{\Gamma}}{{\rm d } x {\rm d } y}$,
derived for the decay $B \to X_u \ell \nu_\ell$ in \cite{greubrey},
which we have checked, after changing the normalization  for 
\bxsll . Defining the
 kinematic variables $(x,y)$ as
$q^2 = x^2 m_b^2$, $v \cdot q = (x+ 1/2 (1-x)^2 y) m_b$,
the Sudakov-improved Dalitz distribution  is given by
\begin{eqnarray}
\label{doubleexpon}
{d^2 {\cal{B}} \over d x d y}(B \to X_s \ell^+ \ell^-) &=& - {\cal{B}}_0 
 \frac{8}{3} x (1 - x^2)^2 (1 + 2 x^2) \,
\exp\Big( - {2 \alpha_s \over 3 \pi} \ln^2 (1 - y) \Big)
  \\
&\times &  \left\{
{4 \alpha_s \over 3 \pi} {\ln(1-y) \over (1-y)}
\Big[ 1 - {2 \alpha_s \over 3 \pi} \big( G(x) + H(y) \big) \Big]
-{2 \alpha_s \over 3 \pi} {d H \over d y}(y) \right\}  C_9^2
 \, ,\nonumber
\end{eqnarray}
where the functions $G(x)$ and $H(y)$ can be seen in \cite{greubrey}. 
The constant ${\cal{B}}_0$ is defined below.

The most significant effect of the bound state is the difference between
$m_B$ and $m_b$ 
which is dominated by $\bar{\Lambda}$. 
The spectrum $\frac{{\rm d}{\cal{B}}}{{\rm d } S_H}$
is obtained  
after changing variables
from $(x,y)$ to $(q^2,S_H)$ and performing an integration over $q^2$.
It is valid in the region
$m_B \frac{m_B\bar{\Lambda}-\bar{\Lambda}^2+m_s^2}{m_B-\bar{\Lambda}}
< S_H \leq m_B^2 $ (or $m_B \bar{\Lambda} \leq S_H \leq m_B^2$, neglecting
$m_s$) which excludes the zeroth order and virtual gluon
kinematics ($s_0=m_s^2$).
The hadronic invariant mass spectrum thus found depends rather
sensitively on $m_b$ (or equivalently $\bar{\Lambda}$).
An analogous analysis for the decay
$B \to X_u \ell \nu_\ell$ has been performed in \cite{FLW}.

The hadronic tensor in eq.~(\ref{eq:hadrontensor}) can be expanded in 
inverse powers of $m_b$ with the help of the 
HQET techniques \cite{FLSold,georgi,MW}.
The leading term in this expansion, i.e., ${\cal O}(m_b^0)$ reproduces the 
parton model result. In HQET, the next to leading power corrections are
parameterized in terms of $\lambda_1$ and $\lambda_2$. 
The contributions of the power corrections
to the structure functions $T_i$ has been calculated up to (but not 
including) $O(1/m_b^3)$ and given in \cite{AHHM97}.
After contracting the hadronic and leptonic tensors and
with the help of the kinematic identities given in eq.~(\ref{eq:kin}),
we can make the dependence on $x_0$ and $\s_0$ explicit,
\begin{eqnarray}
        {T^{L/R}}_{\mu \nu} \, {L^{L/R}}^{\mu \nu} = 
                {m_b}^2 \left\{ 2  (1-2 x_0+\s_0)  {T_1}^{L/R} 
                + \left[ x_0^2 - \frac{1}{4} \u^2 - \s_0 \right] {T_2}^{L/R} 
                \mp  (1-2 x_0+\s_0) \u \, {T_3}^{L/R} \right\} 
        \nonumber
\end{eqnarray}
and with this we are able to derive the double differential power corrected 
spectrum $\frac{{\rm d} {\cal{B}}}{{\rm d} x_0 \, {\rm d}\s_0}$ for
\bxsll. 
Integrating eq.~(\ref{eqn:dgds}) over $\u$ first, where 
the variable $\u$ is bounded by 
$-2 \sqrt{x_0^2-\s_0} \leq \u \leq +2\sqrt{x_0^2-\s_0}$,
we arrive at the following expression \cite{AH98-2}: 
\begin{eqnarray}
\frac{{\rm d}^2 {\cal{B}}}{{\rm d} x_0 \, {\rm d}\s_0}& =&-\frac{8}{\pi} 
{\cal B}_{0}
{\mbox{Im}}\sqrt{x_0^2-\s_0}
\left\{ (1-2 x_0+\s_0)T_1(\s_0,x_0)+\frac{x_0^2-\s_0}{3}T_2(\s_0,x_0) \right\}
+ {\cal{O}}(\lambda_i \alpha_s)
\label{doublediff} \, .
\end{eqnarray}
As the structure function $T_3$ does not
contribute to
the branching ratio, we did not consider it in our present work.
The functions $T_1(\s_0,x_0)$ and $T_2(\s_0,x_0)$ have been derived by
us after a lengthy calculation and the resulting expressions are too
long to be given here. They can be seen together with other details of
the calculations in \cite{AH98-2}.

The branching ratio for \bxsll is usually expressed in terms
of the measured semileptonic branching ratio ${\cal B}_{sl}$
for the decay $B \to X_c \ell \nu_\ell$. This fixes
the normalization constant ${\cal B}_0$ to be,
\begin{equation}
        {\cal B}_0 \equiv
                {\cal B}_{sl} \frac{3 \, \alpha^2}{16 \pi^2} \frac{
    {\vert V_{ts}^* V_{tb}\vert}^2}{\absvcb^2} \frac{1}{f(\mc) \kappa(\mc)}
                \; ,
\label{eqn:seminorm}
\end{equation}
where $f(\mc)$
is the phase space factor for $\Gamma (B \rightarrow X_c \ell \nu_{\ell})$
and
the function $\kappa(\mc)$ accounts for both the $O(\alpha_s)$ QCD 
correction to 
the semileptonic decay  width \cite{CM78} and the leading order
$(1/m_b)^2$ power correction \cite{georgi}. They are given explicitly in 
\cite{FLS}.

The hadron energy spectrum can now be obtained by integrating over $\s_0$.
The kinematic boundaries are given as:
$max(\ms^2,-1+2 x_0 +4 \ml^2) \leq \s_0 \leq x_0^2$, 
$\ms \leq   x_0   \leq \frac{1}{2} (1+\ms^2-4 \ml^2)$.
Here we keep $\ml$ as a regulator wherever it is necessary.
Including the leading power corrections, the  
hadron energy spectrum in the decay  \bxsll is derived by us and given in
\cite{AH98-2}.

The lowest spectral moments
in the decay \bxsll  at the parton 
level are worked out by taking into account the
two types of corrections discussed earlier, namely the leading power $1/m_b$ 
and the perturbative ${\cal{O}}(\alpha_s)$ corrections.
To that end, we define:
\begin{equation}
{\cal M}^{(n,m)}_{l^{+} l^{-}} \equiv 
   {1\over {\cal B}_0}\int (\hat s_0-\hat m_s^2)^n  x_0^m\,
   {{\rm d} {\cal B}\over{\rm d}\hat s_0{\rm d} x_0}
   \,{\rm d}\hat s_0{\rm d} x_0\,,
\end{equation}
for integers $n$ and $m$.  These moments are related to the 
corresponding moments $\langle x_0^m(\hat s_0-\hat m_s^2)^n\rangle$  
obtained at the parton level 
by a scaling factor which yields the corrected branching
ratio ${\cal B}={\cal B}_0 {\cal M}_{\ell^+ \ell^-}^{(n,m)}$.
Thus, 
   $\langle x_0^m(\hat s_0-\hat m_s^2)^n\rangle =
{{\cal B}_0\over {\cal B}}\,
   {\cal M}^{(n,m)}_{l^{+} l^{-}}$.
We remind that one has to Taylor expand the correction factor ${\cal 
B}_0/{\cal B}$ in terms of the ${\cal O}(\alpha_s)$ and power corrections. 
The moments can be expressed as double expansion in ${\cal{O}}(\alpha_s)$
and $1/m_b$ and to the accuracy of our calculations they can be 
represented in the following form:
 \begin{eqnarray}
 {\cal M}^{(n,m)}_{l^{+} l^{-}}=D_0^{(n,m)}+
\frac{\alpha_s}{\pi} {C_9}^2 A^{(n,m)}+
\lo D_1^{(n,m)} + \lt D_2^{(n,m)} \,\, ,
\end{eqnarray}
with a further decomposition into pieces from different Wilson 
coefficients for $i=0,1,2$:
\begin{eqnarray}
\label{momentexp}
D_i^{(n,m)}=\alpha_i^{(n,m)} {C_7^{{\mbox{eff}}}}^2+
\beta_i^{(n,m)} C_{10}^2+
\gamma_i^{(n,m)} C_7^{{\mbox{eff}}} +\delta_i^{(n,m)}.
\end{eqnarray}
The terms $\gamma_i^{(n,m)}$ and $\delta_i^{(n,m)}$ in              
eq.~(\ref{momentexp}) result from the terms proportional 
to ${\it{Re}}(C_9^{{\mbox{eff}}})C_7^{{\mbox{eff}}}$ and
$|C_9^{{\mbox{eff}}}|^2$  in  eq.~(\ref{doublediff}), respectively.   
The explicit expressions for   
$\alpha_i^{(n,m)},\beta_i^{(n,m)},  \gamma_i^{(n,m)}, \delta_i^{(n,m)}$
are given in \cite{AH98-2}.

The leading perturbative contributions for the hadronic invariant mass and 
hadron energy 
moments can be obtained analytically,  
\begin{eqnarray}
A^{(0,0)}&=&\frac{25-4 \pi^2}{9} \, ,
~~A^{(1,0)}=\frac{91}{675} \, ,
~~A^{(2,0)}=\frac{5}{486} \, ,\nonumber \\
A^{(0,1)}&=&\frac{1381-210 \pi^2}{1350} \, ,
A^{(0,2)}=\frac{2257-320 \pi^2}{5400} \, .
\label{eq:A10}
\end{eqnarray}
The zeroth moment $n=m=0$ is needed for the normalization and we recall 
that the result for $A^{(0,0)}$ was derived by Cabibbo and Maiani 
in the context of the ${\cal O}(\alpha_s)$ correction to the semileptonic
decay rate $B \to X \ell \nu_\ell$ quite 
some time ago \cite{CM78}.
Likewise, the first mixed moment $A^{(1,1)}$ can be extracted from 
the results given  
in \cite{FLS} for the decay $B \to X \ell \nu_{\ell}$ after changing the 
normalization, $A^{(1,1)}=3/50$.
For the lowest order parton model contribution 
$D_0^{(n,m)}$, we find, in agreement 
with \cite{FLS}, that the first two hadronic invariant mass moments 
$\langle \s_0-\ms^2 \rangle, \, \langle(\s_0-\ms^2)^2 \rangle$ and the first 
mixed moment $\langle x_0 (\s_0-\ms^2) \rangle$ 
vanish: $D_0^{(n,0)}=0$, for $n=1,2$ and $D_0^{(1,1)}=0 $ .

We can eliminate the hidden dependence on the non-perturbative parameters 
resulting from the $b$-quark mass in the moments 
${\cal M}^{(n,m)}_{l^{+} l^{-}}$ with the help of the HQET mass relation.
As $m_s$ is of order $\Lambda_{QCD}$, to be consistent we keep only 
terms up to order $m_s^2/m_b^2$ \cite{FLSphenom}. An additional 
$m_b$-dependence is in the mass ratios $\ml=\frac{m_l}{m_b}$.
With this we obtain the moments for the 
physical quantities valid up to ${\cal{O}}(\alpha_s/m_B^2,1/m_B^3)$,
where the second equation corresponds to a further use of
$m_s={\cal{O}}(\Lambda_{QCD})$.
We get for the first two hadronic invariant mass moments
{\footnote{Our first expression for 
$\langle S_H^2\rangle$, eq.~(\ref{sHmoments}), 
does not agree in the coefficient of 
$\langle\hat s_0-\hat m_s^2\rangle $ with the one given in \cite{FLS} 
(their eq.~(4.1)). We point out that $m_B^2$ should have been replaced by 
$m_b^2$ in this expression.
This has been confirmed by Adam Falk (private communication).
Dropping the higher order terms given in their expressions, 
the hadronic moments in HQET derived here and in
 \cite{FLS} agree.}}
\begin{eqnarray}\label{sHmoments}
   \langle S_H\rangle&=&m_s^2+\bar\Lambda^2+(m_B^2-2\bar\Lambda m_B
)\,\langle\hat s_0-\hat m_s^2\rangle
+(2\bar\Lambda m_B-2\bar\Lambda^2-\lambda_1-3\lambda_2)
  \langle  x_0\rangle\,, \nonumber\\
  \langle S_H^2\rangle&=&
m_s^4+2\bar\Lambda^2 m_s^2+
  2 m_s^2 (m_B^2-2\bar\Lambda m_B)
  \langle\hat s_0-\hat m_s^2\rangle
+2m_s^2 (2\bar\Lambda m_B-2\bar\Lambda^2-\lambda_1-3\lambda_2)
\langle  x_0\rangle
  \nonumber\\
  &&\quad\mbox{}+ 
(m_B^4-4\bar\Lambda m_B^3
)\langle (\hat s_0-\hat m_s^2)^2\rangle
+ 4\bar\Lambda^2 m_B^2 \langle x_0^2\rangle+
  4\bar\Lambda m_B^3\langle x_0(\hat s_0-\hat m_s^2)\rangle\,,  \\
&=&
(m_B^4-4\bar\Lambda m_B^3)\langle (\hat s_0-\hat m_s^2)^2\rangle
+ 4\bar\Lambda^2 m_B^2 \langle x_0^2\rangle+
  4\bar\Lambda m_B^3\langle x_0(\hat s_0-\hat m_s^2)\rangle \,,
\nonumber
\end{eqnarray}
and for the hadron energy moments:
\begin{eqnarray}\label{EHmoments}
  \langle E_H\rangle &=& \bar\Lambda-{\lambda_1+3\lambda_2\over2m_B}
  +\left(m_B-\bar\Lambda+{\lambda_1+3\lambda_2\over2m_B}\right)\langle
   x_0\rangle\,,\nonumber\\
  \langle E_H^2\rangle &=& \bar\Lambda^2 + (2\bar\Lambda m_B -
2\bar\Lambda^2
  -\lambda_1-3\lambda_2)\langle  x_0\rangle\\
  &&\quad +(m_B^2-2\bar\Lambda m_B+\bar\Lambda^2+\lambda_1+3\lambda_2)
  \langle x_0^2\rangle\,.\nonumber
\end{eqnarray}
One sees that
there are linear power corrections, ${\cal O}(\bar{\Lambda}/m_B)$,
present in all these hadronic quantities except 
$\langle S_H^2 \rangle$ which starts in
$\frac{\alpha_s}{\pi} \frac{\bar{\Lambda}}{m_B}$.

Using the expressions for the HQET moments derived by us \cite{AH98-2}, 
we  present the numerical results for the hadronic moments in \bxsll.
The parameters used in arriving at the numerical coefficients are given in 
Table~\ref{wilson} and specified below:
\begin{eqnarray}
\label{eq:param}
m_W &=& 80.26 ~\mbox{(GeV)}, ~~~m_Z=91.19 ~\mbox{(GeV)}, 
~~~\sin^2 \theta_W =0.2325~, \nonumber\\
m_s &=& 0.2 ~\mbox{(GeV)},
~~~m_c = 1.4 ~\mbox{(GeV)}, 
~~~m_b = 4.8 ~\mbox{(GeV)}, 
~~~m_t = 175 \pm 5 ~\mbox{(GeV)}, \nonumber  \\
\mu &=&{m_{b}}^{+m_{b}}_{-m_{b}/2},     
~~~\alpha^{-1} = 129 ,   
~~~\alpha_s (m_Z) = 0.117 \pm 0.005 ~, 
~~~{\cal B}_{sl} = (10.4 \pm 0.4)\% .    
\end{eqnarray}
Inserting the expressions for the moments calculated at the partonic
level into 
eqs.~(\ref{sHmoments}) and (\ref{EHmoments}), 
we find the following expressions for the short-distance hadronic moments, 
valid up to ${\cal{O}}(\alpha_s/m_B^2,1/m_B^3)$:
\begin{eqnarray}
   \langle S_H\rangle&=&m_B^2 (\frac{m_s^2}{m_B^2}
+0.093 \frac{\alpha_s}{\pi} 
-0.069 \frac{\bar{\Lambda}}{m_B} \frac{\alpha_s}{\pi}
+0.735 \frac{\bar{\Lambda}}{m_B}+0.243 \frac{\bar{\Lambda}^2}{m_B^2}
+ 0.273 \frac{\lambda_1}{m_B^2}-0.513\frac{\lambda_2}{m_B^2}) \nonumber \, ,\\
\label{eq:hadmoments}
 \langle S_H^2\rangle&=&m_B^4 (0.0071 \frac{\alpha_s}{\pi} 
+0.138 \frac{\bar{\Lambda}}{m_B} \frac{\alpha_s}{\pi}
+0.587\frac{\bar{\Lambda}^2}{m_B^2}
-0.196 \frac{\lambda_1}{m_B^2}) \, ,\\
   \langle E_H\rangle&=& 0.367 m_B  (1+0.148 \frac{\alpha_s}{\pi} 
-0.352 \frac{\bar{\Lambda}}{m_B} \frac{\alpha_s}{\pi}
+1.691 \frac{\bar{\Lambda}}{m_B}+0.012\frac{\bar{\Lambda}^2}{m_B^2}
+ 0.024 \frac{\lambda_1}{m_B^2}+1.070\frac{\lambda_2}{m_B^2}) \, ,\nonumber \\
 \langle E_H^2\rangle&=&0.147 m_B^2 (1+0.324 \frac{\alpha_s}{\pi} 
-0.128 \frac{\bar{\Lambda}}{m_B} \frac{\alpha_s}{\pi}
+2.954 \frac{\bar{\Lambda}}{m_B}+2.740\frac{\bar{\Lambda}^2}{m_B^2}
-0.299 \frac{\lambda_1}{m_B^2}+0.162\frac{\lambda_2}{m_B^2}) \, .\nonumber
\end{eqnarray}
Concerning the non-perturbative parts related to the
$c\bar{c}$ loop in \bxsll, it has been suggested
in \cite{buchallaisidorirey} that an
${\cal{O}}(\Lambda^2_{QCD}/m_c^2)$ expansion in the context of HQET
can be carried out to
take into account such effects  in
the invariant mass spectrum away from the resonances.
Using the expressions (obtained with $m_s=0$)
for the $1/m_c^2$ amplitude, we have calculated the
partonic energy moments
$\triangle \langle x_0^n \rangle$,
which correct the short-distance result at order $\lambda_2/m_c^2$:
\begin{eqnarray}
\triangle \langle x_0^n \rangle {{\cal B}\over {\cal B}_0}
&=&-\frac{256 C_2 \lambda_2}{27 m_c^2}
\int_0^{1/2(1-4 \ml^2)} dx_0 x_0^{n+2} {\rm Re} \left[
F(r) \left( C_9^{\mbox{eff}} (3-2 x_0)+2 C_7^{{\mbox{eff}}}
\frac{-3+4 x_0+2 x_0^2}{2 x_0-1} \right) \right] \; ,
\nonumber 
\end{eqnarray}
where $r=(1-2 x_0)/4 \mc^2$ and $F(r)$ is given in 
\cite{buchallaisidorirey}. 
The invariant mass and mixed moments give zero contribution in the order we
are working for $m_s=0$.
Thus, the correction to the hadronic mass moments are vanishing, if we
further neglect terms proportional to
$\frac{\lambda_2}{m_c^2} \bar{\Lambda}$ and $ \frac{\lambda_2}{m_c^2}
\lambda_i$, with $i=1,2$.
For the hadron energy moments we obtain numerically
\begin{eqnarray}
\triangle \langle E_H \rangle_{1/m_c^2}&=&
m_B \triangle \langle x_0 \rangle= -0.007 \, {\mbox{GeV}} \; ,
\nonumber \\
\triangle \langle E_H^2 \rangle_{1/m_c^2}&=&
m_B^2 \triangle \langle x_0^2 \rangle= -0.013 \, {\mbox{GeV}}^2 \; ,   
\end{eqnarray}
leading to a correction of order $-0.3 \%$
to the short-distance values presented in Table \ref{tab:emoments}.

\begin{figure}[htb]
\vskip -0.5truein
\centerline{\epsfysize=3.5in
{\epsffile{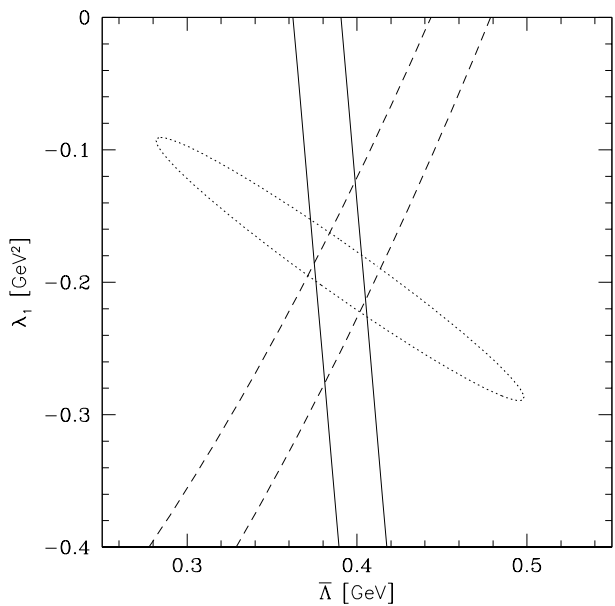}}}
\vskip -0.3truein
\caption[]{ \it $\langle S_H \rangle$ (solid bands) and  $\langle S_H^2 
\rangle$ (dashed bands)
correlation in ($\lambda_1$-$\bar{\Lambda}$) space for  
$\langle S_H \rangle =1.64\pm 0.06$ GeV $^2$ and $\langle S_H^2 \rangle =4.48
\pm 0.29$ 
GeV $^4$, corresponding to the values in Table \ref{tab:emoments}.
The curves are forced to meet at the point 
$\lambda_1=-0.2$ GeV$^2$ and $\bar{\Lambda}=0.39$ GeV. The correlation
from the analysis of the decay $B
\to X \ell \nu_\ell$ from ref.~\cite{gremm} is also shown here (ellipse).}
\label{fig:laml1}
\end{figure}

With the help of the expressions given above,
we have calculated numerically the hadronic moments in HQET for the decay 
$B \to X_s \ell^{+} \ell^{-}$, $\ell=\mu,e$ and have estimated the errors
by varying the parameters within their $\pm 1 \sigma$ ranges given in 
eq.~(\ref{eq:param}). They are presented in Table {\ref{tab:emoments}}
where we have used $\bar{\Lambda}=0.39 \, {\mbox{GeV}}$ and 
$\lambda_1=-0.2 \, {\mbox{GeV}}^2$.
Further, using $\alpha_s(m_b)=0.21$ and $\lambda_2=0.12 \, {\mbox{GeV}}^2$,
the explicit dependencies of the hadronic moments given in 
eq.~(\ref{eq:hadmoments}) on the HQET parameters
$\lambda_1$ and $\bar{\Lambda}$ can be worked out. 
\begin{eqnarray}
   \langle S_H\rangle&=&0.0055 m_B^2(1+
132.61 \frac{\bar{\Lambda}}{m_B}+44.14 \frac{\bar{\Lambda}^2}{m_B^2}
+ 49.66 \frac{\lambda_1}{m_B^2}) \nonumber \, ,\\
 \langle S_H^2\rangle&=& 0.00048 m_B^4(1+
19.41 \frac{\bar{\Lambda}}{m_B} 
+1223.41\frac{\bar{\Lambda}^2}{m_B^2}
-408.39 \frac{\lambda_1}{m_B^2}) \, .
\end{eqnarray}
As expected, the
dependence of the energy moments $\langle E_H^n\rangle$ on $\bar{\Lambda}$
and $\lambda_1$ is very weak and we do not show these here.
While interpreting these numbers, one should bear in mind that there are two
comparable expansion parameters $\bar{\Lambda}/m_B$ and $\alpha_s/\pi$ and we
have fixed the latter in showing the numbers.
  The correlations on the HQET parameters $\lambda_1$ and $\bar{\Lambda}$
 which follow from (assumed) fixed
values of the hadronic invariant mass moments  $\langle S_H \rangle$ 
 and  $\langle S_H^2 \rangle$ are shown in Fig.~\ref{fig:laml1}. We
have taken the values for the decay $B \to X_s \mu^+ \mu^-$ from Table 
\ref{tab:emoments} 
for the sake of illustration and have also shown the presently irreducible
theoretical errors on these moments following from the input parameters 
$m_t$, $\alpha_s$
and the scale $\mu$, given in eq.~(\ref{eq:param}). The errors were 
calculated 
by varying these parameters in the indicated range, one at a time,
and adding the individual errors in quadrature. As the entries in Table 
\ref{tab:emoments} are calculated for 
the best-fit values of $\lambda_1$ and $\bar{\Lambda}$ taken from  the 
analysis of Gremm et al. \cite{gremm} for
the electron energy spectrum in $B \to X \ell \nu_\ell$, there is no
surprise that these curves meet at this point. This exercise
has to be repeated with real data in \bxsll to draw any 
quantitative conclusions. Using the CLEO cuts on hadronic and dileptonic
masses \cite{cleobsll97}, we estimate that $O(200)$ \bxsll ($\ell=e,\mu$) 
events will be available per $10^7$ $B\bar{B}$ hadrons \cite{AH98-2}.
So, there will be plenty of \bxsll decays in the forthcoming B facilities
to measure the correlation shown in Fig.~\ref{fig:laml1}. 
 
The theoretical stability of the moments has to be checked against
higher order corrections and the 
error estimates presented here will have to be improved.  
 The ``BLM-enhanced" two-loop corrections
\cite{BLM} proportional to $\alpha_s^2\beta_0$, where $\beta_0 = 11 -2 
n_f/3$ is the first term in the QCD beta function, can be included
at the parton level as has been done in other decays \cite{FLS,GS97}, but 
not being crucial to our point we have not 
done this. More importantly, higher order corrections in 
$\alpha_s$ and $1/m_b^3$ are not included here.
 While we do not think that
the higher orders in $\alpha_s$ will have a significant influence, the
second moment $\langle S_H^2 \rangle$ is susceptible to the presence of
$1/m_b^3$ corrections as shown for the decay $B \to X
\ell \nu_\ell$ \cite{FL98}. This will considerably enlarge the theoretical
error represented by the dashed band for $\langle S_H^2 \rangle$ in
Fig.~\ref{fig:laml1}. Fortunately, the coefficient of the 
$\bar{\Lambda}/m_B$ term in $\langle S_H \rangle$
is large. Hence, a good measurement of  this moment alone constrains 
$\bar{\Lambda}$ effectively. Of course, the utility 
of the hadronic moments calculated above is only in conjunction
with the experimental cuts. Since 
the optimal experimental cuts in \bxsll remain to be defined, we hope to
return to this and related issue of doing an improved
theoretical error estimate in a future publication. The power corrections 
presented here in the hadron spectrum and 
hadronic spectral moments in \bxsll are the first results in this decay.
\begin{table}[h]
        \begin{center}
        \begin{tabular}{|c|l|l|l|l|}
        \hline
        \multicolumn{1}{|c|}{HQET}      & 
                \multicolumn{1}{|c|}{$\langle S_H\rangle$  } &  
\multicolumn{1}{|c|}{$\langle S_H^2\rangle$ } &
                \multicolumn{1}{|c|}{$\langle E_H\rangle$  } &  
\multicolumn{1}{|c|}{$\langle E_H^2\rangle$ } \\
 \hline
\multicolumn{1}{|c|}{{\mbox{}}} &
\multicolumn{1}{|c|}{$({\mbox{GeV}}^2)$ } &
\multicolumn{1}{|c|}{$({\mbox{GeV}}^4)$ } &
\multicolumn{1}{|c|}{$({\mbox{GeV}})$ } &
\multicolumn{1}{|c|}{$({\mbox{GeV}}^2)$ } \\
 \hline
$\mu^+ \mu^-$&$1.64 \pm 0.06$ &$4.48 \pm 0.29$ & $2.21 \pm 0.04 $&
$5.14 \pm 0.16$ \\
$e^+ e^-$   &$1.79 \pm 0.07 $ &$4.98 \pm 0.29$ & $2.41 \pm 0.06 $& 
$6.09 \pm 0.29$\\
        \hline   
        \end{tabular}
        \end{center}
\caption{\it Hadronic spectral moments for $B \to X_s \mu^{+} \mu^{-}$
and $B \to X_s e^{+} e^{-}$ 
in HQET with $\bar{\Lambda}=0.39 \, GeV$ and $\lambda_1=-0.2 \, GeV^2$.
The quoted error results from varying $\mu, \alpha_s$ and the 
top mass within the ranges given in eq.~(\ref{eq:param}).}
\label{tab:emoments}
\end{table}
In summary, we have calculated the ${\cal O}(\alpha_s)$
 perturbative 
QCD and leading ${\cal O}(1/m_b)$ corrections to the hadron spectra in the 
decay \bxsll, including the Sudakov-improvements in the
perturbative part. Hadronic invariant mass spectrum is calculable
in HQET over a limited range $S_H > m_B \bar{\Lambda}$ and it depends 
sensitively on the parameter $\bar{\Lambda}$ (equivalently $m_b$). 
These features are qualitatively very
similar to the ones found for the hadronic invariant mass spectrum in 
the decay $B \to X_u \ell \nu_\ell$ \cite{FLW}.
The $1/m_b$-corrections to the parton model hadron energy spectrum in 
\bxsll are small over most part of this spectrum. However, heavy quark
expansion breaks down near the low end-point of this spectrum and near the
$c\bar{c}$ threshold.
We have calculated the spectral hadronic moments $\langle E_H^n \rangle$
and $\langle S_H^n \rangle$ for $n=1,2$ and have worked out their dependence
on the HQET parameters $\bar{\Lambda}$ and $\lambda_1$.
The correlations in \bxsll are shown to be different
than the ones in the semileptonic decay $B \to X \ell \nu_\ell$. This
allows, in principle, a method to determine them from data in \bxsll.
We show the kind of constraints following from a Gedanken experiment
in \bxsll and the present analysis of data in $B \to X \ell 
\nu_\ell$ \cite{gremm} to illustrate this point. 
  
\bigskip
\noindent

 We thank Christoph Greub for helpful discussions.
 Correspondence with Adam Falk and Gino Isidori on
power corrections are thankfully acknowledged.
\end{document}